\documentclass[conference]{IEEEtran}
\IEEEoverridecommandlockouts
\usepackage{cite}
\usepackage{amsmath,amssymb,amsfonts}
\usepackage{algorithmic}
\usepackage{graphicx}
\usepackage{textcomp}
\usepackage{xcolor}
\usepackage{graphicx}
\usepackage{tikz}
\usetikzlibrary{matrix}
\usepackage{cite}
\usepackage{amsmath,amssymb,amsfonts}
\usepackage{algorithmic}
\usepackage{textcomp}
\usepackage{hyperref}
\usepackage{booktabs}
\usepackage{xspace}
\usepackage{svg}
\usepackage{amsmath}
\usepackage{footnote}
\usepackage{makecell}
\usepackage{multirow}
\usepackage{enumitem}
\usepackage{array}
\usepackage[nolist,nohyperlinks]{acronym}
\usepackage{hyperref}
\usepackage{subcaption}
\usepackage{caption, subcaption}
\usepackage[acronym]{glossaries}
\acrodef{AI}{Artificial Intelligence}
\acrodef{GI}{Gastrointestinal}
\acrodef{ML}{Machine Learning}
\acrodef{DL}{Deep learning}
\acrodef{CNN}{Convolutional Neural Network}
\acrodef{CADx}{computer aided diagnosis} 
\acrodef{FPS}{frame per second}
\acrodef{FDA}{Food and Drug Administration}
\acrodef{GDPR}{General Data Protection Regulation}
\acrodef{HIPAA}{Health Insurance Portability and Accountability Act}
\acrodef{CAD}{Computer-Aided diagnosis}

\begin{document}

\title{Vision Transformer for Efficient Chest X-ray and Gastrointestinal Image Classification}

\author{\IEEEauthorblockN{Smriti Regmi\IEEEauthorrefmark{1},
Aliza Subedi\IEEEauthorrefmark{1}, 
Ulas Bagci\IEEEauthorrefmark{2},
Debesh Jha\IEEEauthorrefmark{2}}\\
 \IEEEauthorrefmark{1}Pashchimanchal Campus, Nepal\\ 
\IEEEauthorrefmark{2} Machine \& Hybrid Intelligence Lab, Department of Radiology, Northwestern University, Chicago, USA} 

\maketitle

\begin{abstract}
Medical image analysis is a hot research topic because of its usefulness in different clinical applications, such as early disease diagnosis and treatment. Convolutional neural networks (CNNs) have become the de-facto standard in medical image analysis tasks because of their ability to learn complex features from the available datasets, which makes them surpass humans in many image-understanding tasks. In addition to CNNs, transformer architectures also have gained popularity for medical image analysis tasks. However, despite progress in the field, there are still potential areas for improvement. This study uses different CNNs and transformer-based methods with a wide range of data augmentation techniques. We evaluated their performance on three medical image datasets from different modalities. We evaluated and compared the performance of the vision transformer model with other state-of-the-art (SOTA) pre-trained CNN networks. For Chest X-ray, our vision transformer model achieved the highest F1 score of 0.9532, recall of 0.9533, Matthews correlation coefficient (MCC) of 0.9259, and ROC-AUC score of 0.97. Similarly, for the Kvasir dataset, we achieved an F1 score of 0.9436, recall of 0.9437, MCC of 0.9360, and  ROC-AUC score of 0.97. For the Kvasir-Capsule (a large-scale VCE dataset), our ViT model achieved a weighted F1-score of 0.7156, recall of 0.7182, MCC of 0.3705, and  ROC-AUC score of 0.57. We found that our transformer-based models were better or more effective than various CNN models for classifying different anatomical structures, findings, and abnormalities.  Our model showed improvement over the CNN-based approaches and suggests that it could be used as a new benchmarking algorithm for algorithm development. 
\end{abstract}


\section{Introduction}
Medical image analysis is crucial for the diagnosis and prognosis of many medical disorders. Timely detection of lesions, tumors, and other anatomical anomalies indicative of life-threatening medical conditions can facilitate early and effective treatment. Traditionally, radiologists and clinicians performed the majority of medical image interpretations. However, their analysis is susceptible to inter-observer variability, fatigue, and a significant miss-rate \cite{quekel1999miss,ahn2012miss}. Thus, clinicians cannot consistently analyze and interpret medical data with high accuracy. Similarly, various disorders, including lung infections, covid-19 (SARc-CoV2), pneumonia, heart disease, and injuries can be diagnosed using chest X-rays. It is essential to get an early and precise diagnosis of these lung disorders for effective treatment. Diseases of the digestive tract, which includes the oesophagus, stomach, small intestine, and colon are referred to as gastrointestinal (GI) disorders. These ailments can vary from benign conditions like acid reflux to potentially fatal conditions like colon cancer. Early detection and accurate diagnosis can allow prompt and effective treatment, which leads to improved patient outcomes and enhanced quality of life. Thus, \ac{CAD} system could assist physicians and clinical experts in making early, accurate, and efficient diagnoses. 

\acp{CNN} has a built-in bias for recognizing patterns in data and their convolution operations can be adapted to fit the specific characteristics of the data. This combination of inductive bias and flexibility enables CNNs to perform exceptionally well in tasks related to computer vision, such as image classification, object detection, and segmentation. Similarly, there are popular methodologies for automated medical image analysis (for example, medical image segmentation \cite{article}, classification \cite{7064414}, registration \cite{8633930}, and reconstruction \cite{article1}. Likewise, ViT \cite{dosovitskiy2020image} is an image classification model built from the architecture of a transformer initially intended for text-based operations. While a transformer accepts a series of tokens as input, ViT accepts image patches instead. It is evident that training ViT with a sufficient amount of data generates impressive outcomes. It outperforms equivalent state-of-the-art CNNs with four times less computational effort. Similarly, the Data-Efficient Image Transformer (DeiT) \cite{touvron2021training} is a modified version of ViT built to reduce the ViTs' data dependency. To train the DeiT model, a teacher-student approach designed for transformers is used. It uses a distillation token to ensure that the student learns from the teacher through attention.

Despite having successful deep learning-based algorithms in the community, there is a need for standard efficient algorithms that could be integrated into the clinical workflow for different medical applications. In this work, we explore and examine several \acp{CNN} and Transformer based approaches for medical image classification tasks. As a use case, we analyze Chest X-ray images associated with different pathologies for classification. Similarly,  we examine the images captured during live endoscopy and colonoscopy examinations and classify anatomical landmarks, pathological findings and abnormalities, and regular findings from \ac{GI} tract. Additionally, we examine the images captured using video capsule endoscopy that comprises different classes of small bowel, bleeding, angiectasia, polyp, and ulcer. For our experiments, we explored several different methodologies. We used DeiT, ViT, and Ensemble models created by stacking pre-trained CNN models and transfer learning method with pre-trained CNN architecture and compared the results between these classification models.

The main contribution of our work is as follows: 
\begin{enumerate}
\item  We have explored and investigated different convolutional neural network-based approaches, along with pre-trained backbone networks, and compared them with the recent Transformer based approach ViT \cite{dosovitskiy2020image} and DeiT \cite{touvron2021training} for medical image classification. We performed extensive experiments on three publicly available medical datasets (endoscopy, video capsule endoscopy, and computed tomography (CT)). 

\item ViT \cite{dosovitskiy2020image} showed state-of-the-art performance on all three datasets for most metrics and surpassed CNN and DeiT-based models. The ViT approach obtained the highest Matthews correlation coefficient (MCC) score for all three medical datasets, which is an important metric for imbalanced datasets. Additionally, we also achieved a promising Receiver operating (ROC) curve score of ViT on all presented datasets.

\end{enumerate}

\section{Related Work}
Our work relates to \ac{CNN}, Vision Transformer, Chest X-ray, and \ac{GI} tract diseases findings. Here, we observe the representative works closely related to these topics.


\subsection{Chest X-ray}
CNNs have recently been utilized in various studies to analyze Chest X-ray  (CXR) images and identify lung diseases \cite{anthimopoulos2016lung,alakwaa2017lung}, including pneumonia \cite{varshni2019pneumonia} and tuberculosis \cite{liu2017tx}.
By the middle of the year 2020, researchers were able to successfully develop CNN-based methodologies for COVID-19 detection from CXR images.  Wang et al.\cite{wang2020covid} proposed COVID-Net algorithm for the detection of COVID-19 cases using CXR images. Additionally, they proposed the COVIDx dataset and leveraged GSInquire \cite{lin2019explanations} to analyze the more discriminative features contributing towards the prediction in order for the COVID-Net to aid clinicians. Tahir et al. \cite{tahir2022deep} used transfer learning of multiple pre-trained CNN models and classified different coronavirus families (SARS, MERS, and COVID-19) with sensitivity values greater than 90\%. Chikkara et al.\cite{chhikara2020deep} investigated the likelihood of diagnosing pneumonia from CXR images and assessed the performance of a few pre-trained models (ResNet, Xception, and Inception) using pre-processing techniques, including filtering and gamma correction.

\subsection{Gastrointestinal disease}
\acp{CNN} have been explored extensively for classification of \ac{GI} diseases and findings. Thambawita et al. \cite{thambawita2018medico} presented five different \ac{ML} and \ac{DL} approaches based on global features and pre-trained CNN with transfer learning. Their CNN-based approaches, along with the transfer learning, outperformed \ac{ML}-based approach. The combination of Resnet-152~\cite{he2016deep} and Densenet-161~\cite{huang2017densely} with an additional MLP achieved the highest performance. Recently, Afriyie et al.~\cite{afriyie2022gastrointestinal} presented a less sophisticated but still effective pre-processing technique for identifying endoscopic images known as denoising capsule networks (Dn-CapsNets). In addition, they created activation maps (AM) utilizing the feature representations to display the outcomes. The trained model obtained 94.16\%, 83.1\%, 86.7\%, 96.1\%, and 86.6\% in terms of accuracy, precision, sensitivity, specificity, and F1-score respectively. Improvements in accuracy have been observed while comparing the proposed method with other SOTA methods. Srivastava et al.~\cite{srivastava2022video} proposed FocalConvNet, a focal modulation network combined with lightweight convolutional layers to classify real-time anatomical and luminal findings (pathological and mucosal view) classification in video capsule endoscopy. They achieved the weighted F1-score, recall, and MCC of 0.6734, 0.6373, and 0.2974 respectively, outperforming SOTA methodologies.

\begin{figure*}[!h]
    \centering
    \includegraphics[width=0.99\textwidth]{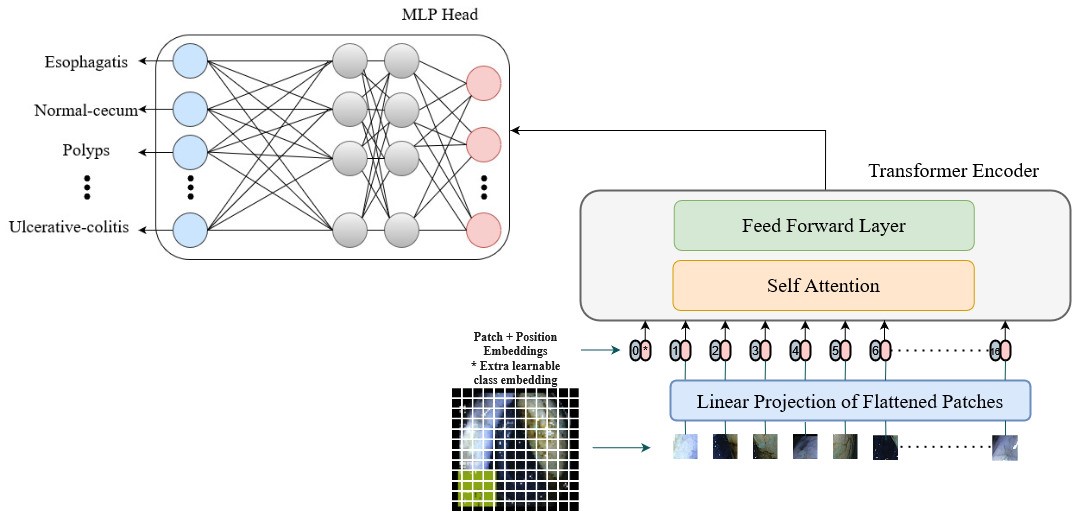}
   
    \caption{An original ViT \cite{dosovitskiy2020image} structure for the classification task. The image is first converted into flattened patches through Patch Embedding and Position Embedding, then processed by the Transformer encoder \cite{vaswani2017attention}. The prediction result is obtained after the MLP Head.}
    \label{fig:my_label}
\end{figure*}

\subsection{Transformer network}

Recently transformer-based network has been successful because of their capacity to model long-range dependencies, global receptive field, and lack of inductive bias in contrast to CNNs. Dosovitskiy et al.\cite{dosovitskiy2020image} states transformers may be used directly for image classification. They have demonstrated that a pure transformer applied directly to a sequence of image patches may successfully perform well on image classification tasks without the reliance on CNNs. Here, the series of image patches are directly fed to the typical transformer encoder \cite{vaswani2017attention}. Similarly, Usman et al. \cite{usman2022analyzing}, Matsoukas et al. \cite{matsoukas2021time} compared the performance of Vision Transformers (ViTs) and CNN models trained on mainstream medical image classification tasks.
Their findings suggest that transfer learning-based ViTs can achieve comparable performance levels as CNNs in analyzing medical datasets. Moreover, Hosain et al.~\cite{hosain2022gastrointestinal} describe a technique for assisting medical diagnosis procedures and identifying gastrointestinal tract disorders using a vision transformer and transfer learning models. In their study, they proposed a vision transformer-based method with a 95.63\% accuracy using wireless capsule endoscopy (WCE) curated images of the colon. They demonstrated that vision transformer outperformed DenseNet201 \cite{huang2017densely}. 
Likewise, Mabrouk et al. \cite{mabrouk2022pneumonia} proposed an Ensemble method for pneumonia classification using the images of  Chest X-rays.

\section{Methodology}
In our study, we utilized various variants of Vision Transformer (ViT) \cite{dosovitskiy2020image} models pre-trained on ImageNet-21k, including ViT-B/16, ViT-L/16, and ViT-L/32. Here, ViT-B/16 means the base variant. Similarly,
ViT-L/16 and ViT-L/32 refer to the large variant with the input patch size of 16×16 and 32×32 respectively. The ViT models were fine-tuned on three medical domain datasets. Two components are transferred via this method: the model architecture and its weights obtained via pretraining on the ImageNet-21k.

\subsection{ViT architecture}

Vision Transformer is a state-of-the-art architecture based on the transformer architecture originally developed for natural language processing tasks. Firstly, the original image  $x \in \mathbb{R}^{H \times W \times C}$ is split into $16\times 16$  non-overlapping patches then flattened to  $1\times n$ matrices. The 2D image $x \in \mathbb{R}^{H \times W \times C}$ is reshaped into the sequence of flattened 2D patches $x_p \in \mathbb{R}^{N \times (P^2 \cdot C)}$ where (H, W ) is the resolution of the original image, C is the number of channels, (P, P ) is the resolution of each image patch, and $N = \frac{HW}{P^2}$ is the resulting number of the patches. These flattened patches are projected to D dimension via a trainable linear projection layer and can be represented in matrix form as  $X \in \mathbb{R}^{N \times D}$.

Then, a positional embedding is added to retain the positional information.
\begin{equation}
    x = \hat{x} + E_{pos}, E_{pos} \in \mathbb{R}^{N \times D}
\end{equation}
After that, the resulting tokens are fed to the transformer encoder consisting of L stacked base blocks. Each base block consists of a multi-head self-attention
and a multi-layer perceptron (MLP), with Layer-Norm (LN).
\begin{equation}
z'_l = \text{MSA}(\text{LN}(z_{l-1})) + z_{l-1}, l = 1,...,L
\end{equation}
\begin{equation}
z_l = \text{MLP}(\text{LN}(z'_l)) + z'_l, l = 1,...,L
\end{equation}

More information about the ViT architecture, self-attention and multi-head attention can be found in~\cite{dosovitskiy2020image}.

\begin{figure}[!h]
\scriptsize
\centering
\includegraphics[width=0.5\textwidth]{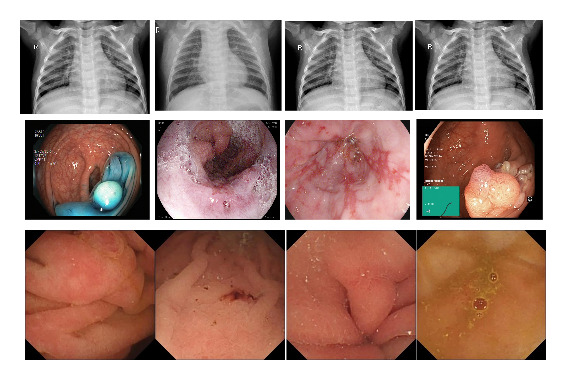}
\caption{Example samples from Chest X-ray, Kvasir, and  Kvasir-Capsule  datasets.}
\label{fig:datasetsample}
\end{figure}

\begin{table*} [!t]
    \centering
 \caption{Summary of the datasets used in our experiments.}
      \label{table:dataset}
    \begin{tabular}{ c| c| c| c| c| c| c} 
    \toprule
\textbf{Dataset} & \textbf{No. of Images} &\textbf{Input size} & \textbf{Train} & \textbf{Valid} & \textbf{Test} &  \textbf{Application}\\ \midrule
\shortstack{Chest X-ray dataset} & 7135 & variable & 5693 & 671 & 771 & Lung disease\\ \hline
 Kvasir dataset  & 8000 & $720\times 556$ & 6,400 & 800  & 800 &Endoscopy\\   \hline
 Kvasir-Capsule & 44,152 & $336\times 336$ & 19,280& 4,820 & 23061 & Bowel disease \\
\bottomrule
\end{tabular}
\end{table*}


\section{Experiments}
\subsection{Datasets}

To evaluate the performance of the model architecture, we used three publicly available medical datasets from different modalities. The sample of the dataset can be observed from Figure~\ref{fig:datasetsample}. We use \textbf{Chest X-ray dataset} \cite{cohen2020covid}~\cite{kermany2018labeled}~\cite{rahman2020reliable}, which contains 7135  chest x-ray images (576 COVID-19, 1583 Normal, 4273 Pneumonia, 703 Tuberculosis) of patients, collected from different publicly available resources as the first dataset. Our second dataset is  \textbf{Kvasir}~\cite{pogorelov2017kvasir}. It is a multi-class dataset consisting of 1,000 images per class with a total of 8,000 images for eight different classes. These classes consist of pathological findings (esophagitis, polyps, ulcerative colitis), anatomical landmarks (z-line, pylorus, cecum), and normal and regular findings (normal colon mucosa, stool), and polyp removal cases (dyed and lifted polyps, dyed resection margins).
The \textbf{Kvasir-Capsule dataset}~\cite{smedsrud2021kvasir} is the third dataset used in our study. Kvasir-Capsule is the largest publicly available video capsule endoscopy dataset comprising 44,228 labelled images from 13 classes of
anatomical and luminal findings. There are different numbers of images available for each class in the dataset. Specifically, there are 1529 images for the Pylorus class, 10 images for the Ampulla of Vater class, and 4189 images for the Ileocecal Valve class. Additionally, there are 34338 images of Normal Clean Mucosa, 2906 images of Reduced Mucosal View, 592 images of Lymphangiectasia, 159 images of Erythema, 866 images of Angiectasia, 446 images of Blood-Fresh, 12 images of Blood-Hematin, 506 images of Erosion, 854 images of Ulcer, 55 images of Polyp, and 776 images of Foreign body. The dataset is heavily imbalanced. Table~\ref{table:dataset} presents more descriptions of the datasets.


\subsection{Implementation details}
\subsubsection{Vision Transformer}
Different designs of ViT models \cite{dosovitskiy2020image} are fine-tuned in our experiments. ViT models, namely ViT-Base variant with $16\times16$ (ViT-B/16) input patch size and ViT-Large variant with $16\times16$ (ViT-L/16) input patch size and $32\times32$ (ViT-L/32) input patch size, are adopted in all of the experiments in this work. ViT-Base consists of 12 transformer blocks, each with a 12-headed self-attention module. There are a total of 86M trainable parameters in the ViT-Base model. Likewise, ViT-Large consists of 24 transformer blocks with a 16-headed self-attention module. There are a total of 307M trainable parameters in the ViT-Large model.

\vspace{0.3mm}

\subsubsection{Pre-Training}
In this study, the implemented architectures were run using the Tensorflow \cite{abadi2016tensorflow} framework. The images from the Chest X-ray and the Kvasir dataset were resized to $224\times224$, while the images from the Kvasir-Capsule dataset were resized to $64\times64$. The DeiT-Ti and DeiT-B 384 models had image resolutions of $224\times224$ and $384\times384$, respectively, for all datasets. The datasets were partitioned into training, validation, and test sets as indicated in Table~\ref{table:dataset}. For Kvasir-Capsule dataset, given~\href{https://github.com/simula/kvasir-capsule/tree/master/official_splits}{official split} (split1) was used.

\vspace{0.2mm}

\subsubsection{Data Augmentation}
Various data augmentation strategies were employed for different datasets in this study to address data constraints. For the Chest X-ray dataset, techniques such as rescaling, adjusting brightness, and random zooming, as well as horizontal and vertical shifting, were applied. For the Kvasir dataset, the images were augmented by altering brightness, rotating, and flipping them vertically. Similar techniques, including image rotation, vertical and horizontal flip, and shift, were used for the Kvasir-Capsule dataset.

\subsection{Models} 
Vision Transformer-based models used in our experiments are the recommended models with different sizes, i.e., ViT-B and ViT-L. In these models, B and L are the representations of base and large models.
Similarly, we fine-tuned different variants of DeiT \cite{touvron2021training}, including DeiT-Ti and DeiT-B on medical datasets. For the comparison of the transformer with CNN models, we use pre-trained CNN networks, including DenseNet201\cite{huang2017densely}, DenseNet121 \cite{huang2017densely}, InceptionResNetV2 \cite{szegedy2017inception}, Xception \cite{chollet2017xception}, MobileNetV2 \cite{sandler2018mobilenetv2} and an ensemble of two models (DenseNet201 and DenseNet121) formed using the soft voting technique.

\subsection{Experiment setup and configuration}
In order to enhance performance on a certain task, we manually tuned the hyperparameters by altering them using trial and error. To get the optimal performance on a particular dataset, many parameters like learning rate, batch size, loss function, optimizer, trainable parameters, and decay rate were modified. For the Chest X-ray and Kvasir datasets, the categorical cross-entropy loss function is used. Given that the Kvasir-Capsule dataset is imbalanced, the focal loss \cite{lin2017focal} is applied in this circumstance. Focal Loss (FL), an advanced version of cross-entropy loss is used here to address the class imbalance issue by giving harder or more easily misclassified examples more weights and deweighting simple examples.
\subsection{Evaluation metrics}
We have used widely accepted computer vision metrics for our image classification tasks that include Matthews correlation coefficient (MCC), Frames Per Second (FPS), weighted average precision, F1 score and recall, and overall accuracy. In multi-class classification problems, the weighted average method adjusts for class imbalance by assigning a weight proportional to the number of instances in each class. The standard deviations for Precision, Recall, and F1-score are provided to show the variability of the results. Furthermore, we performed a paired t-test to compare the MCC achieved by the best performing model (in terms of MCC) to the MCC achieved by other state-of-the-art methods. The paired t-test results include the reported p-values.
We have also plotted  the ROC curve and compared the performance of models for each dataset. Additionally, a confusion matrix is plotted for each class, showing the correct and incorrect findings of all three different modality datasets.

\begin{table*}[!t]
\scriptsize
\centering
\caption{Quantitative results on Chest X-ray\cite{cohen2020covid,kermany2018labeled,rahman2020reliable} dataset.}
\begin{tabular}{l|l|l|l|l|l|l|l}
\toprule
 \textbf{Method} & \textbf{Precision} &\textbf{Recall} & \textbf{F1-score} & \textbf{Accuracy}   & \textbf{MCC}& \textbf{P-values}& \textbf{FPS}\\ \midrule
 
DenseNet201 + DenseNet121 & 0.9345 ± 0.0324 & 0.9326 ± 0.0584 & 0.9319 ± 0.0236 & 0.9326 &  0.8931 & 1.24e-03 & 16.08\\\hline
 
DenseNet201  &  0.9169 ± 0.0358 &  0.9157 ± 0.0538  & 0.9150 ± 0.0195  &0.9157 & 0.8658 & 8.41e-05 & 22.47 \\ \hline 

DenseNet121   &  0.9314 ± 0.0358 &  0.9287 ± 0.0678  & 0.9277 ± 0.0370  & 0.9287 & 0.8878 &3.37e-04 & 26.15\\ \hline 
 
InceptionResNetV2 & 0.9415 ± 0.0358   &   0.9403 ± 0.0678  & 0.9400 ± 0.0370 & 0.9287 &  0.9052  & 1.18e-04 & 20.48\\ \hline 
Xception  & 0.9380 ± 0.0324  &  0.9377 ± 0.0461 &  0.9375 ± 0.0358 &  0.9377 & 0.9010 & 4.80e-05 & 21.34 \\  \hline
MobileNetV2 & 0.9133± 0.0625 &  0.9092 ± 0.0508 &  0.9099 ± 0.0401 &  0.9092 & 0.8567 & 8.96e-04 & 21.85\\ \hline

DeiT-Ti  &   0.9266 ± 0.0453 &  0.9261 ± 0.0326 &  0.9262 ± 0.0368 &  0.9261 & 0.8821 & 3.37e-04 & 25.48 \\ \hline
 DeiT-B 384  &  0.9332 ± 0.0592 &  0.9300 ± 0.0728 &  0.9291 ± 0.0272 &  0.9300 & 0.8899 & 1.10e-03 & 16.75 \\ \hline
 ViT-L/32  &  \textcolor{red}{0.9527} ± 0.0362  & \textcolor{red}{0.9520} ± 0.0301  & \textcolor{red} {0.9521} ± 0.0277 & \textcolor{red}{0.9520} &  \textcolor{red}{0.9236} & 3.83e-02 & {11.49} \\  
 \hline
 ViT-L/16  &   \textbf{0.9533} ± 0.0202  & \textbf{0.9533} ± 0.0301  &  \textbf{0.9532} ± 0.0225 & \textbf{0.9533} &  \textbf{0.9259} & - & {11.20} \\  \hline
  ViT-B/16  & 0.9291 ± 0.0592  & 0.9248 ± 0.0728  & 0.9248 ± 0.0272 & 0.9248 & 0.8813  & 3.77e-04 & 20.74 \\ 
   \bottomrule
  \end{tabular}
\label{table: Result chest xray}
\end{table*}

\begin{table*}[!t]
\scriptsize
\centering
\caption{Quantitative results on the Kvasir~\cite{pogorelov2017kvasir} dataset.}
 \label{table:KvasirCapsule}
 \begin{tabular}{c|c|c|c|c|c|c|c}
 \toprule
 \textbf{Method} & \textbf{Precision} &\textbf{Recall} & \textbf{F1-score} & \textbf{Accuracy}   & \textbf{MCC}& \textbf{P-values}& \textbf{FPS}\\ \midrule

 DenseNet201 + DenseNet121 & 0.9284 ± 0.0559 & 0.9263 ± 0.0711 &0.9258 ± 0.0538 & 0.9263 &  0.9161 & 2.60e-01 & 12.32\\\hline
 
 DenseNet201  & 0.9262 ± 0.0467&  0.9250 ± 0.0628  & 0.9246 ± 0.0472 & 0.9250  &  0.9146 & 1.52e-01 & 21.59 \\ \hline 

DenseNet121  &  0.9136 ± 0.0523 &  0.9112 ± 0.0749  & 0.9105 ± 0.0515   &0.9113 & 0.8991 & 6.63e-02 & 24.35 \\ \hline 
 
InceptionResNetV2 & 0.8877 ± 0.0619   &   0.8875 ± 0.0753  & 0.8870 ± 0.0648 & 0.8875 &  0.8716  & 1.90e-02 &21.18 \\ \hline 
 Xception &  0.9032 ± 0.0611 & 0.9025 ± 0.0761 & 0.9020 ± 0.0639 & 0.9025  & 0.8888 & 4.64e-02 & 43.12  \\  \hline

MobileNetV2  & 0.8789 ± 0.0689  &  0.8775 ± 0.0826 &  0.8769 ± 0.0691 &  0.8775 
 & 0.8603 & 1.42e-02 &43.97\\ \hline

 DeiT-Ti  & 0.9353 ± 0.0337 & 0.9350 ± 0.0394
 & 0.9349 ± 0.0338  & 0.9350 &0.9258 & 5.82e-01 & 25.80 \\  \hline

 DeiT-B 384 & 0.9408 ± 0.0401 & \textcolor{red} {0.9401} ± 0.0466 & \textcolor{red}{0.9399
 }± 0.0370  & \textcolor{red}{0.9401}  & \textcolor{red}{0.9319} & 3.73e-01 & 15.81\\  \hline

ViT-L/32  &0.9375 ± 0.0532  & 0.9337 ± 0.0698 & 0.9333 ± 0.0458 & 0.9337 & 0.9249 & 4.55e-01 & 13.35 \\  
 \hline

 ViT-L/16 &   \textbf{0.9454 }± 0.0400  & \textbf{0.9437} ± 0.0487  &  \textbf{0.9436} ± 0.0345 & \textbf{0.9437} &  \textbf{0.9360} & - & 11.34\\  \hline

ViT-B/16   & \textcolor{red}{0.9433} ± 0.0427  & 0.9400 ± 0.0532  & 0.9398 ± 0.0260 & 0.9400 & 0.9316 & 5.53e-01 & 21.50 \\ \hline

\end{tabular}
\label{table:Kvasir result}
\end{table*}

\begin{table*}[!t]
\scriptsize
\centering
\caption{Quantitative results on the Kvasir-Capsule \cite{smedsrud2021kvasir} dataset.}
\begin{tabular}{c|c|c|c|c|c|c|c}
 \toprule
 \textbf{Method} & \textbf{Precision} &\textbf{Recall} & \textbf{F1-score} & \textbf{Accuracy}   & \textbf{MCC}& \textbf{P-values}& \textbf{FPS}\\ \midrule

 DenseNet201 + DenseNet121 & \textcolor{red}{0.6633} ± 0.2485 & \textbf{0.7230} ± 0.2859 & \textcolor{red}{0.6737} ± 0.2594 & \textbf{0.7230} &  0.3560 & 9.63e-03 & 578.15\\\hline 
 
 DenseNet201 &0.6606 ± 0.2522 & \textcolor{red}{0.7198} ± 0.2820  & 0.6737 ± 0.2583
  & \textcolor{red}{0.7198 }  & \textcolor{red}{0.3585} & 8.88e-03 & 1142.90\\ \hline 
  
  DenseNet121 &  0.6564 ± 0.2660 &  0.7187 ± 0.2848  & 0.6720 ± 0.2621   & 0.7187 & 0.3500 &4.28e-03 &1143.35   \\ \hline 
 
 InceptionResNetV2 & 0.6059 ± 0.2647   &   0.6900 ± 0.2742  & 0.6548 ± 0.0336 & 0.6203 &  0.2277  & 3.06e-04 & 544.81 \\ \hline 
 Xception & 0.6159 ± 0.2484  &  0.6895  ± 0.2710 &  0.6310 ± 0.2431 &  0.6895 & 0.2544 &3.34e-04 & 1223.44 \\  \hline

 MobileNetV2 & 0.5903 ± 0.2645   &   0.6800 ± 0.2732  & 0.5925 ± 0.2342 & 0.6800 &  0.1637 & 7.041e-04& 3555.12 \\ \hline 

 DeiT-Ti & 0.6100 ± 0.2603 & 0.6839 ± 0.2669 & 0.6203 ± 0.2431   & 0.6839  & 0.2212 & 5.06e-05 & 281.24 \\  \hline
 DeiT-B 384  & 0.6496 ± 0.3020 & 0.7180 ± 0.2770  & 0.6657 ± 0.3010  & 0.7185   & 0.3422 & 6.10e-04 & 520.21\\  \hline
 ViT-L/32  &   0.6483 ± 0.2922  & 0.7182 ± 0.2985 &  0.6631 ± 0.2760 & 0.7182 &  0.3377 & 3.11e-02 & 343.60 \\  
 \hline
 ViT-L/16  &  0.6425 ± 0.2806  & 0.6751  ± 0.2725  &  0.6405 ± 0.2581 & 0.6751 &  0.2637 & 1.69e-03 & 262.09\\  \hline
 ViT-B/16   & \textbf{0.6841} ± 0.2985  & 0.7156  ±0.2899   & \textbf{0.7156}  ± 0.2779
 & 0.7156 & \textbf{0.3705}  & - & 570.53 \\ 
 \hline
  \end{tabular}
\label{table:kvasir capsule result}
\end{table*}

\section{Results}
The performance of the trained model was evaluated using various quantitative indicators. The results for the Chest X-ray dataset are shown in Table~\ref{table: Result chest xray}. Similarly, the results for the Kvasir dataset are shown in Table~\ref{table:Kvasir result}. From Table~\ref{table: Result chest xray}, it can be observed that ViT-L/16 obtained an MCC of 0.9259, which is nearly 2\% more than the pre-trained CNNs and DeiT-based models. ViT-L/16 also obtains the highest score in all other metrics except macro average precision. For the Kvasir dataset, ViT-L/16 obtains the highest MCC of 0.9360. The most competitive model to our model is DeiT-B 384, which obtains 0.38\% less than the best score.  Likewise, Table~\ref{table:kvasir capsule result} presents the quantitative results for the Kvasir-Capsule dataset. It shows that the ViT-B/16 obtains the highest MCC score of 0.3705. The most competitive MCC to ViT-based approach was DeiT-B 384. Additionally, our model also outperformed other models in terms of weighted precision, and F1-score. DenseNet201 and its ensemble show better results for some metrics such as weighted recall and accuracy. 

\begin{figure}[!t]
\includegraphics[trim=0.3cm 0.55cm 0cm 1.5cm, clip, width=1\linewidth, height=7.45cm]{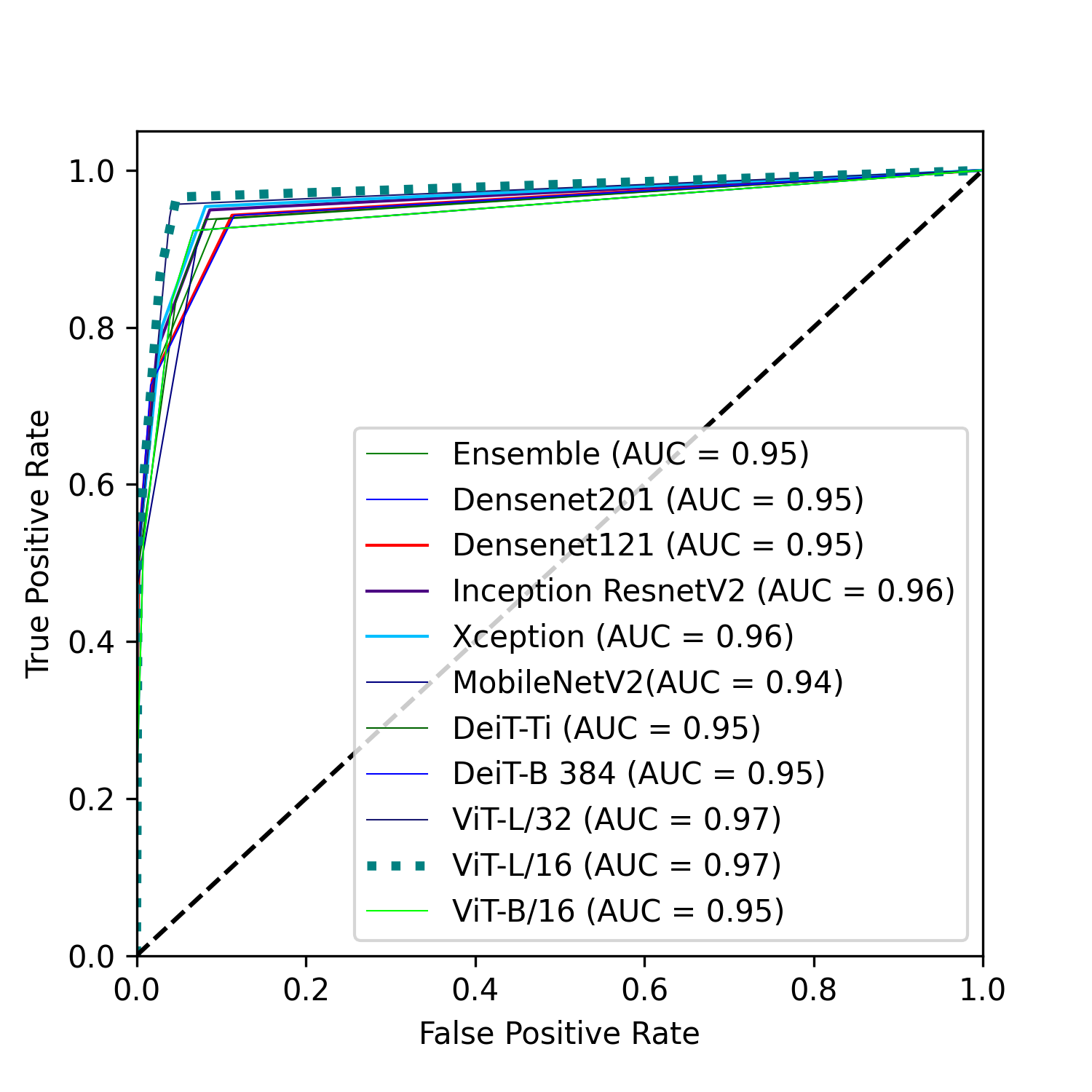}
\caption{ROC curve for Chest X-ray dataset.}
\label{fig:a6}
\end{figure}

\begin{figure}[!t]
\includegraphics[trim=0.3cm 0.55cm 0cm 0.2cm, clip, width=1\linewidth, height=7.5cm]{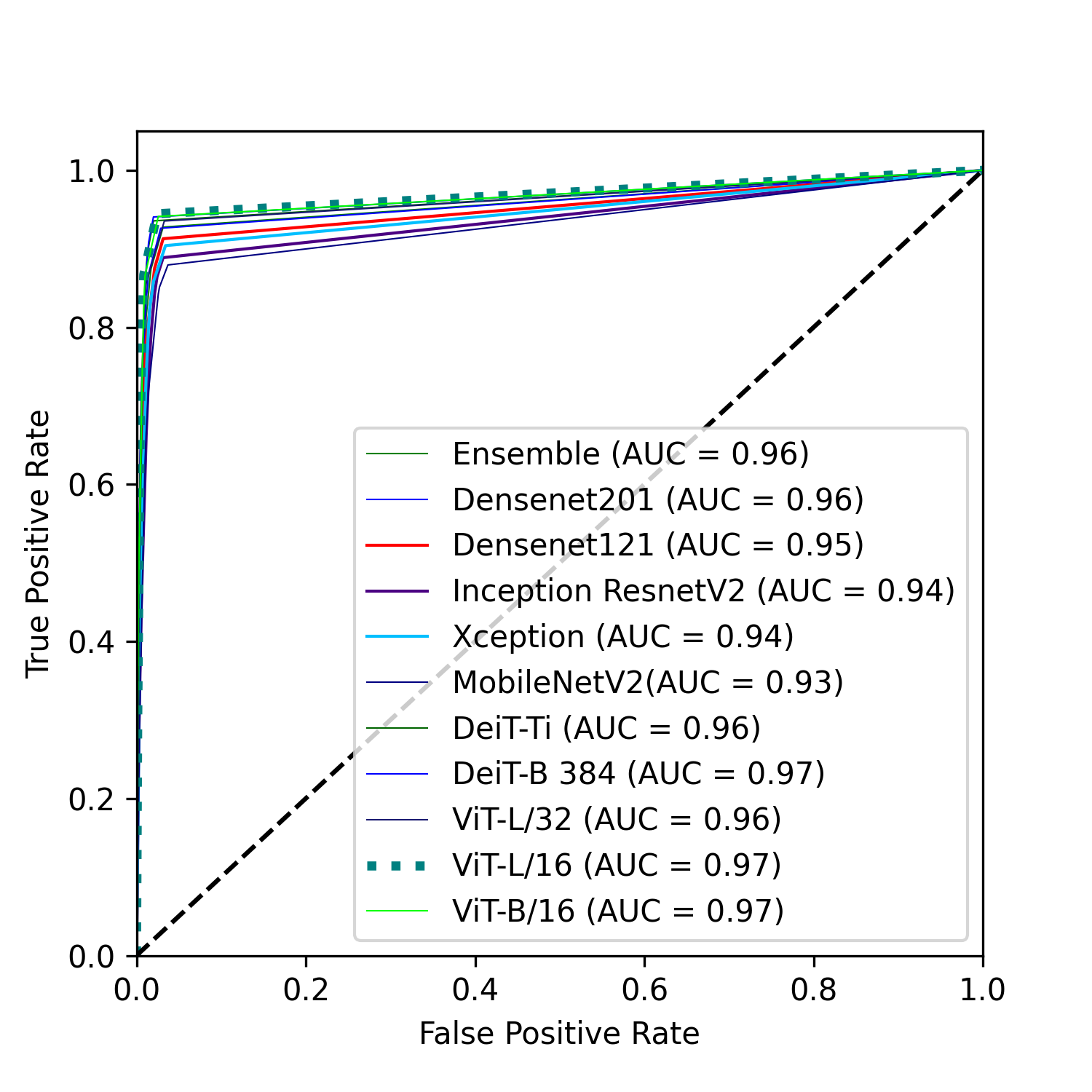} 
\caption{ROC curve for Kvasir dataset.}
\label{fig:a7}
\end{figure}

\begin{figure}[!t]
\includegraphics[trim=0.3cm 0.55cm 0cm 0.2cm, clip, width=1\linewidth, height=7.41cm]{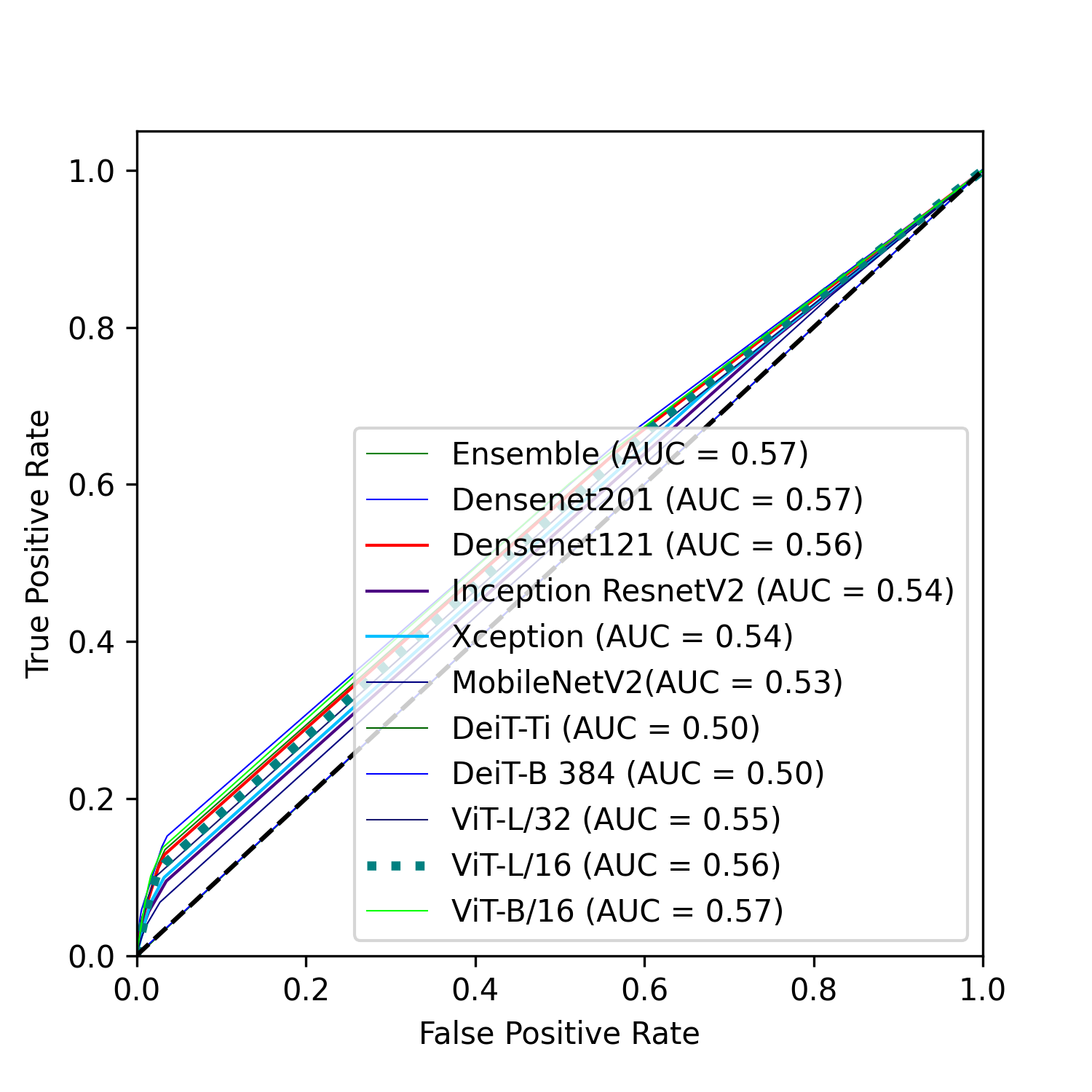}
\caption{ROC curve for Kvasir-Capsule dataset.}
\label{fig:a8}
\end{figure}

Table \ref{table: Result chest xray},~\ref{table:Kvasir result},~\ref{table:kvasir capsule result} shows the p-values obtained in our paired t-test experiments. The p-values indicate whether the difference between the MCC of our model and the MCC of each SOTA method is statistically significant, with a p-value less than 0.05 being typically considered significant. For the chest X-ray dataset and Kvasir capsule dataset, all models' p-value is less than 0.05, indicating that the difference between the best-performing ViT model and other SOTA methods is statistically significant. However, for the Kvasir dataset, some of the models have p-values higher than 0.05, indicating that the differences between ViT-L/16 and these methods not being statistically significant.
Furthermore, we generated ROC curves for each model and presented them together in a single graph (Figure \ref{fig:a6},~\ref{fig:a7},~\ref{fig:a8}) to facilitate a performance comparison. Here, the ROC curve of the ViT models shows a promising result for all datasets.


\vspace{2mm}






\section{Discussion}
\vspace{-1mm}
Our research involved fine-tuning different pre-trained models and evaluating their performance on various datasets with varying data modalities. The results in Tables ~\ref{table: Result chest xray},~\ref{table:Kvasir result} indicate that the transformer architecture outperformed other CNN pre-trained models for all metrics. Referring to Table \ref{table:kvasir capsule result}, the results indicate that in the kvasir-capsule dataset, Vision Transformers (ViTs) outperformed CNNs in terms of weighted precision, F1-score, and MCC metrics. This suggests that the attention mechanism of the transformer model enables it to learn more intricate patterns in the data, resulting in more accurate predictions of the images. Additionally, the sequential processing of data in visual transformers makes it possible to use parameters more efficiently, consider longer-term dependencies, and have more generalized learning across different inputs. Furthermore, the absence of recurrence in transformers allows for better generalization and reduces the chances of overfitting.

\subsection{Limitations and open challenges}

Despite the contributions of our study, there are several
limitations that needs to be acknowledged. First off, there are known problems with the labeling quality of the publicly available datasets. While we took steps to mitigate the impact of labeling issues on the results, such as conducting manual verification of a sample of the data, the potential for mislabeled data could still exist. Similarly, we encountered data limitations and resource utility issues. The imbalanced datasets were another challenge in the study. By utilizing several data augmentation approaches, we were able to tackle the data restriction problem.  We hope to resolve our infrastructure issue shortly and look forward to exploring ways to improve the accuracy of medical classification with advanced attention-based Transformer models.


\vspace{2mm}
\subsection{Analysis of the failing cases}
\begin{figure}[!t]
\includegraphics[width=0.7\textwidth,height=4cm]{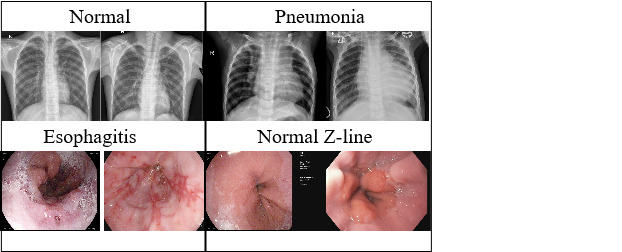} 
\caption{Example frames from  `pneumonia' and `normal' and `esophagitis' and `normal z-line' classes.}
\label{fig:challengesqualitative}
\end{figure}

\begin{figure}[t!]
 \includegraphics[trim=2cm 2.5cm 0cm 0cm, clip, width=0.99\linewidth, height=6.2cm]{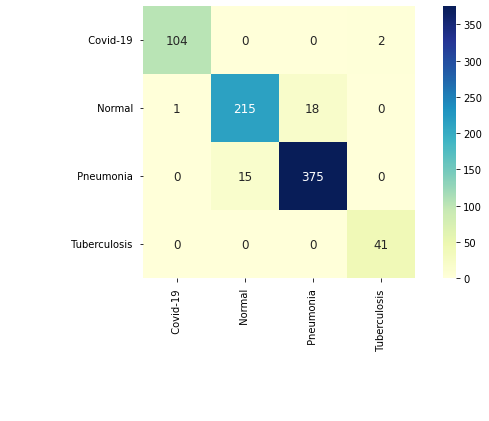}
\caption{Confusion matrix of ViT-L/16 for Chest X-ray dataset.}
\label{fig:a3}
\end{figure}

\begin{figure}[ht!]
\includegraphics[trim=1cm 1.7cm 0cm 0cm, clip, width=0.99\linewidth, height=7.1cm]{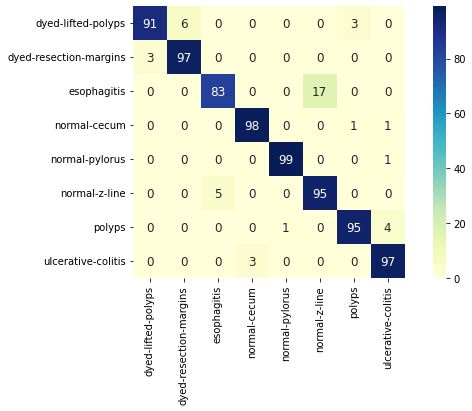} 
\caption{Confusion matrix of ViT-L/16 for Kvasir dataset.}
\label{fig:a4}
\end{figure}

\begin{figure}[t!]
\includegraphics[trim=0.2cm 0.3cm 0cm 0cm, clip, width=0.99\linewidth, height=6.9cm]{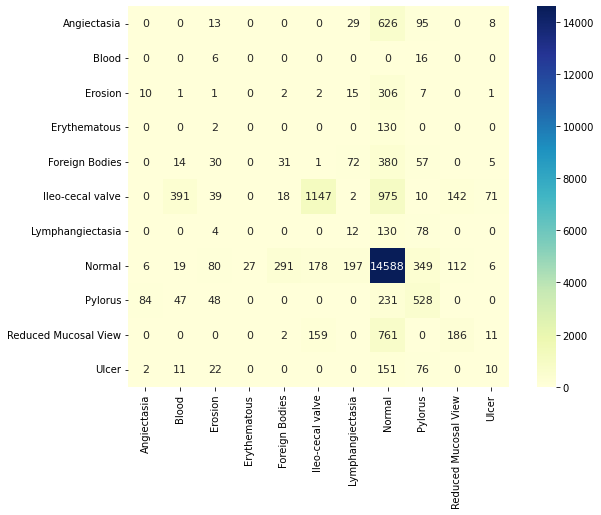}
\caption{Confusion matrix of ViT-B/16 for Kvasir-Capsule dataset.}
\label{fig:a5}
\end{figure}

We further analyzed failed images of all three datasets. Figure~\ref{fig:challengesqualitative} shows some examples of challenging images that led to a performance drop. The most significant misclassification for the Chest X-ray dataset was observed between the `normal' and `pneumonia' classes. The results shown in Figure~\ref{fig:a3} indicate the confusion matrix for the ViT classifier for the Chest X-ray dataset. The primary reason for the confusion is the similar appearance in the chest X-ray images for both classes, making it difficult for the model to differentiate between them. The other reason can be the unequal data distribution in both classes and simple data augmentation.
The primary reason for the confusion is the similar appearance in the chest X-ray images for both classes, making it difficult for the model to differentiate between them. The other reason can be the unequal data distribution in both classes. 
\\
Similarly, for the Kvasir dataset, the model is highly confused between the esophagitis class and the normal z-line class which is clear from Figure~\ref{fig:a4}. 
The main reason for the confusion is the similarity between these classes in terms of their color properties and the overlapping features between these classes. In the image of the normal z-line, the line defining the border between the lower esophagus and the stomach may resemble the same appearance as images of esophagitis, which is an inflammation of the esophagus lining. Because of this, it could be challenging for the model to distinguish between the two classes using only visual features. The other reasons can be model architecture and the choice of the loss function.  
The results from the confusion matrix of the Kvasir-Capsule (Figure~\ref{fig:a5}) indicate that the dataset is highly scattered. The model best classified the ``Normal" category, with 14588 out of 15853 images accurately identified. On the other hand, the categories of \textbf{\textit{``Angiectasia," ``Blood," and ``Erythematous"}} were the most frequently misclassified. The confusion is primarily due to the uneven distribution of data among each class.

\section{Conclusion}
In our study, we compared the classification performance of several pre-trained CNN models, their ensemble models, and transformer-based models (i.e., ViT and DeiT) in three medical domain data sets. Vision transformer achieved MCC values of 0.9259, 0.9360, and 0.3705  for the Chest X-ray, Kvasir, and Kvasir-Capsule datasets. Furthermore, we also calculated p-values for different models with respect to the best model and compared the values with a pre-determined level of significance. The classification performance shows that the attention-based Vision transformer achieved the best classification result, outperforming different CNN models and their ensembles. Several metrics, such as the confusion matrix and ROC curve were used in the study to evaluate the ViT models. This study provides evidence that a vision transformer is a powerful tool for addressing challenging medical classification problems and could be a valuable benchmark for developing deep learning-based classification algorithms. Likewise, this study opens new avenues for applying self-attention-based architectures as an alternative to CNNs in various image classification tasks.

\subsection*{Acknowledgement}
This project is supported by the NIH funding: R01-CA246704 and R01-CA240639.

\bibliographystyle{IEEEtran}
\bibliography{references}

\end{document}